# Complexity of Frictional Interfaces: A Complex Network Perspective

H. O. Ghaffari[1]

*Department of Civil Engineering and Lassonde Institute, University of Toronto, Canada*

M. Sharifzadeh

Faculty of Engineering, Kyushu University, Hakozaki, Japan

E. Evgin

*Department of Civil Engineering, University of Ottawa, Ontario, Canada*

**Abstract:** The shear strength and stick-slip behavior of a rough rock joint are analyzed using the complex network approach. We develop a network approach on correlation patterns of void spaces of an evolvable rough fracture (crack type II). Correlation among networks properties with the hydro -mechanical attributes (obtained from experimental tests) of fracture before and after slip is the direct result of the revealed non-contacts networks. Joint distribution of locally and globally filtered correlation gives a close relation to the contact zones attachment-detachment sequences through the evolution of shear strength of the rock joint. Especially spread of node's degree rate to spread of clustering coefficient rate yielded possible stick and slip sequences during the displacements. Our method can be developed to investigate the complexity

[1] h.o.ghaffari@gmail.com;

hamed.owladeghaffari@utoronto.ca

Dept. Civil Engineering, U of T, MB108-170 College Street-Toronto-ON-Canada

of stick-slip behavior of faults as well as energy /stress localization on crumpled shells/sheets in which ridge networks are controlling the energy distribution.

## 1. Introduction

During the last decade, complex networks have been used increasingly in different fields of science and technology [1-3]. Initial applications of complex networks in geosciences were mostly related to earthquakes [4-6]. Characterization of spatial and temporal structural complexity of such recursive events has been the main objective of the related research [7-13]. Understanding of spatio-temporal topological complexity of events based on field measurements can disclose some other facets of these intra/extra woven events.

Studies pertaining to the topological complexity and its application in some geoscience fields reveals that acquisition and gathering of direct information (especially in temporal scale) is difficult and in many cases are (were) impossible (at least with current technologies). In addition to complex earthquake networks, recently the analysis of climate networks, volcanic networks, river networks and highway networks, as the large scale measurements, have been taken into account [9-13]. In small scales, topological complexity has been evaluated in relation to geoscience fields such as the gradation of soil particles, fracture networks, aperture of fractures, and granular materials [14-20]. The initial step refers to organizational step which tries to find out possible dominant well-known structures within the system. Next step in the most of the mentioned works is to provide a suitable and simple method to yield a similar structure. Such algorithm may support the evolution of structure in spatial or/and temporal cases [21].

May be the most important structural complexity in geological fields is related to fracture networks. Fracture networks with dilatancy [22], joint networks in excavation damaged zones, cracking in pavements (or other natural/man-made structures) and fault networks in large scale have been recognized [23-25]. In the analysis of these networks, the characterization of fractures in a proper space such as friction-displacement space is an essential step. Furthermore, with taking the direct relationship between void spaces and contact areas in to account, one may interest in considering the induced topological complexity of the opening elements (non-frictional contacts) into the fracture behavior. Using linear elastic fracture mechanics, we know aperture or aspect ratio is generally the index to available energy in growth of rupture. Crack like behavior of rupture in frictional interfaces also support the role of contact areas and equivalently apertures. In addition, the variations of fluid flow features (such as permeability and tortuosity) directly are controlled with aperture spaces. In order to characterize the main attributes of the fractured systems, e.g. mechanical and hydraulic properties, several methods have been suggested in the literature [26-30]. Recently, the authors have proposed the implementation of a complex network analysis for the evolution of micro-scale apertures in a rough rock fracture [18-19]. Based on a Euclidean measure, the results confirmed the dependency of hydro-mechanical properties to the attributes of characterized aperture networks. The present study is also related to the complex aperture networks. However, the current study presents the analysis of frictional forces during shearing based on the *correlation* of apertures in a rock joint. The analysis is associated with set up a network on an attribute (such as aperture distribution) in an area. The aforementioned method has also been employed in the analysis of the coupled partial differential equations which was related to two-phase flow [31].

With respect to avalanche-like behavior of collective motion of the ensemble discrete contacts (in the vicinity of a phase transition step), we try to characterize the collective behavior of aperture strings using networks. In this paper we will answer the following three questions: 1) Is there any (hidden) complex structure in the experimentally observed apertures? 2) What is the effect of specific structural complexity of apertures on mechanical response of a fracture? 3) How do apertures regulate with each other to show well-known slip-friction curve? In other words, can we relate the topological complexity of apertures to the evolution path of the fracture?

The organization of the paper is as follows: Section 2 includes a brief description of networks and their characterization. In addition, the construction procedure of aperture networks is explained. Section 3 covers a summary of the experimental procedure. The last section presents the evaluation of the pre- and post-peak (stick-slip) behavior of a rock joint which is followed by the analysis of the constructed network.

## 2. Network of Evolving Apertures

In this section we describe a general method of setting up a network on a fracture surface while the surface property is a superposition of very narrow profiles (ribbons) of one attribute of the system. In other words, one attribute of the system is "granulated "over strings (profiles or ribbons). The relationship between the discrete strings –inferred from long range correlation or elastic forces- results an interwoven network, i.e., topological complexity of interactions. The frictional behavior including the stick-slip response of a joint is related to the sum of real contact areas, which fluctuates with the changes in apertures. It also occurs based on the collective motion and spatially coupled of contact zones. It is shown that the structural complexity of the dynamic aperture changes is controlling and regulating the joint behavior and its unstable response. In order to explain the details of our work, we need to characterize the topological complexity.

A network consists of nodes and edges connecting the [32]. To set up a nondirected network, we considered each string of measured aperture as a node. Each aperture string has $N$ pixels where each pixel shows the void size of that cell. Depending on the direction of strings, the length of the profiles varies. The maximum numbers of strings (in our cases) are in the perpendicular direction to the shear, while the minimum one is in the parallel direction. To make an edge between two nodes, a correlation measurement ( $C_{ij}$ ) over the aperture profiles was used. The main point in the selection of each space is to explore the explicit or implicit hidden relations among different distributed elements of a system. For each pair of signals (profiles) $V_i$ and $V_j$ , containing $N$ elements (pixels) the correlation coefficient can be written as [33]:

$$C_{ij} = \frac{\sum_{k=1}^{N}\left[V_i(k) - \prec V_i \succ\right].\left[V_j(k) - \prec V_j \succ\right]}{\sqrt{\sum_{k=1}^{N}\left[V_i(k) - \prec V_i \succ\right]^2}.\sqrt{\sum_{k=1}^{N}\left[V_j(k) - \prec V_j \succ\right]^2}} \quad (1)$$

where $\prec V_i \succ = \frac{\sum_{k=1}^{N} V_i(k)}{N}$. Obviously, It should be noted that $C_{ij}$ is restricted to $-1 \leq C_{ij} \leq 1$, where $C_{ij}$=1, 0, and -1 are related to perfect correlations, no correlations and perfect anti-correlations, respectively.

Selection of a threshold ( $\xi$ ) to make an edge, can be seen from different views. Choosing a constant value may be associated with the current accuracy of accumulated data where after a maximum threshold the system loses its dominant order. In fact, there is not any unique way in the selection of a constant value, however, preservation of the general pattern of evolution must be considered while the hidden patterns can be related to the several characters of the network. These characters can express different facets of the relations, connectivity, assortivity (hubness), centrality, grouping and other properties of nodes and/or edges [34-36]. Generally, it seems obtaining stable patterns of evolution (not absolute) over a variation of $\xi$ can give a suitable and reasonably formed network [33]. Also, different approaches have been used such as density of links, the dominant correlation among nodes, *c*-*k* space and distribution of edges or clusters. In this study, we set $C_{ij} \geq \xi = 0.2C_{ij}^{\max}$. Considering with this definition, we are filtering uncorrelated profiles over the metric space. In the previous study, the sensitivity of the observed patterns (associated with the Euclidean distance of profiles) has been distinguished [18].

The clustering coefficient describes the degree to which *k* neighbors of a particular node are connected to each other. What we mean by neighbors is the connected nodes to a particular node. The clustering coefficient shows the collaboration between the connected nodes. Assume that the $i^{th}$ node to have $k_i$ neighboring nodes. There can exist at most $k_i(k_i - 1)/2$ edges between the neighbors. We define $c_i$ as the ratio

$$c_i = \frac{actual \quad number \quad of \quad edges \quad between \quad the \quad neighbors \quad of \quad the \quad i^{th} \quad node}{k_i(k_i - 1)/2} \tag{2}$$

Then, the clustering coefficient is given by the average of $c_i$ over all the nodes in the network [21]:

$$C = \frac{1}{N}\sum_{i=1}^{N} c_i . \qquad (3)$$

For $k_i \leq 1$ we define $C \equiv 0$. The closer $C$ is to one the larger is the interconnectedness of the network. The connectivity distribution (or degree distribution), $P(k)$, is the probability of finding nodes with $k$ edges in a network. In large networks, there will always be some fluctuations in the degree distribution. The large fluctuations from the average value ($< k >$) refers to the highly heterogeneous networks while homogeneous networks display low fluctuations [21]. From another perspective, clustering in networks is closely related to degree correlations. Vertex degree correlations are the measures of the statistical dependence of the degrees of neighbouring nodes in a network [35]. Two-point correlation is the criterion in complex networks as it can be related to network assortativity.

The concept of two-point correlation can be included within the conditional probability distribution $P(k' | k)$ that a node of degree $k$ is connected to a node of degree $k'$. In other words, the degrees of neighbouring nodes are not independent. The meaning of degree correlation can also be defined by the average degree of nearest neighbours ($\prec k_{nn} \succ_k$). If $\prec k_{nn} \succ_k$ increases with $k$ high degree nodes (hubs) tend to make a link to high degree nodes, otherwise, if $\prec k_{nn} \succ_k$ decreases with $k$, high degree nodes (hubs) tend to make a link with low degree nodes (disassortative ) [34-36]. From the point of view of fractal complex networks [37-38], the degree correlation may be used as a tool to distinguish the self-similarity of network structures. In fact, in fractal networks large degree nodes (hubs) tend to connect to small degree nodes and not to each other (fractality and disassortativity). Also, the clustering nature of a network can be drawn as the average over all nodes of degree $k$ giving a clustering distribution (or spectrum). In many real-world networks such as the internet the clustering spectrum is a decreasing function of degree which may be interpreted as the hierarchical structures in a network. In contrast, some other networks such social networks and scientific collaborations (and also we will see complex aperture networks) are showing assortative behaviour [35]. It will be shown that spreading of crack like behaviour due to shearing a fracture, can be followed with the patterns of proper spectrum. Similarly, by using the degree correlation, one may define the virtual weight of an edge as an average number of edges connected to the nodes [39].

The average (characteristic) path length $L$ is the mean length of the shortest paths connecting any two nodes on the graph. The shortest path between a pair $(i, j)$ of nodes in a network can be assumed as their geodesic distance, $g_{ij}$, with a mean geodesic distance $L$ given as below [2, 21]:

$$L = \frac{2}{N(N-1)} \sum_{i<j} g_{ij}, \quad (4)$$

where $g_{ij}$ is the geodesic distance (shortest distance) between node $i$ and $j$, and $N$ is the number of nodes. We will use a well known algorithm in finding the shortest paths presented by Dijkstra [40]. Based on the mentioned characteristics of networks two lower and upper bounds of networks can be recognized: regular networks and random networks (or Erdős–Rényi networks [41]). Regular networks have a high clustering coefficient (C $\approx$ 3/4) and a long average path length. Random networks (construction based on random connection of nodes) have a low clustering coefficient and the shortest possible average path length. However Watts and Strogatz [42] introduced a new type of networks with high clustering coefficient and small (much smaller than the regular ones) average path length. This is called small world property.

## 3. Summary of Laboratory Tests

To study the small world properties of rock joints, the results of several laboratory tests were used. The joint geometery consisting of two joint surfaces and the aperture between these two surfaces were measured. The shear and flow tests were performed later on. The rock was granite with a unit weight of 25.9 $kN/m^3$ and uniaxial compressive strength of 172 MPa. An artificial rock joint was made at mid height of the specimen by splitting and using special joint creating apparatus, which has two horizontal jacks and a vertical jack [43-44]. The sides of the joint are cut down after creating the joint. The final size of the sample is 180 mm in length, 100 mm in width and 80 mm in height. Using special mechanical units, various mechanical parameters of this sample were measured. A virtual mesh having a square element size of 0.2 mm was spread on each surface and the height at each position was measured by a laser scanner. The details of the procedure can be found in [45-46]. Different cases of the normal stress (1, 3and 5 MPa) were used while the variation of surfaces were recorded. Figure1 shows the shear strength evolution under different normal loads. In this study, we focus on the patterns, obtained from the test with a 3 MPa normal stress.

## 4. Implementation and Analysis of Complex Aperture Networks

In this section we set up the designated complex network over the aperture profiles, which are perpendicular to the shear direction. By using the correlation measure, the distribution of correlation values along profiles and during the successive shear displacements were obtained (Fig.2). Plotting the correlation distribution shows the transition from a near Poisson distribution to a Gaussian distribution. The change in the type of distribution is followed by the phenomena of the tailing, which is inducing the homogeneity of the correlation values towards high and anti-correlation values. In other words, tailing procedure is tied with the quasi-stable (residual part) states of the joint. Thus, this can be described by reducing the entropy of the system where the clusters of information over correlation space are formed. From another point of view, with considering the correlation patterns, it can be inferred that throughout the shear procedure, there is a relatively high correlation between each profile and the profiles at a certain neighborhood radius. This radius of correlation is increasing non-uniformly (anisotropic development) during shear displacement (Fig.3).

By using the method described in the previous section, a complex aperture network is developed from the correlation patterns (Fig.4). As it can be seen in this figure, the formation of highly correlated nodes (clusters) is distinguishable near the peak point. It can be estimated that the controlling factor in the evolution path of the system is related to the formation of cliques (communities). We will show locality properties of the clusters (intera structures) are much more discriminated at last displacements rather than initial time steps while global variations of the structures are more sensitive to reduction in the shear stress. In fact, forming hubs in the constructed networks may give the key element of synchronization of aperture profiles (or collective motion of discrete contact zones) along the shear process. In other words, reaching to one or multiple attractors and the rate of this reaching after peak point are organized by the spreading and stabilizing the clusters. Unfortunately, due to a low rate of data sampling, the exact evolution of patterns before peak-point is not possible. However, during the discussion on the joint degree correlations, a general concept will be proposed.

The three well-known characteristics of the constructed networks, namely total degree of nodes, clustering coefficient and mean shortest path length are depicted as a function of shear displacement in Figure 5.As it can be followed there is a nearly monotonic growth/decay of the parameters. A considerable sharp change in transition from shear displacement 1 to 2 mm is observed for all three illustrated parameters. This transition is assumed as state transition from the pre-peak to post peak state, while with taking into account the rate of the variation of the parameters the transformation step is discriminated. Also, despite of clustering coefficient trend which show a fully-growth shape the number of edges and mean short length after a shear displacement of 12 mm roughly exhibiting a quasi-stable trend. These results provide the necessary information for the classification of the aperture networks in our rock joint. The high clustering coefficient and low average (characteristic) path length clearly show that our aperture networks have small-world properties.

The development of shear stress over the networks is much faster after the peak point than the pre-peak states. This feature can be explained by understanding the concept of the net-contact areas [59]. At interlocking of asperities step-before maximum static friction- the two-point correlation shows a relatively more uniform shape rather than former and later cases. Also, the current configuration implies that the homogeneity of the revealed network where the nodes with high degree are tending to absorb nodes with low edges. This indicates the property of self-similarity within the network structures. The shear displacements immediately after or near peak (Figure 6) point destroy the homogeneity of the network and spreading slow fronts and dropping of the frictional coefficient is accompanied with a trial to make stable cliques, inducing the heterogeneity to the network structures. Using a microscopic analysis, it can be proven that, for homogenous topologies, many small clusters spread over the network and merge together to form a giant synchronized cluster [54-56]. This event is predicted before reaching to the peak threshold. In heterogeneous graphs, however, one or more central cores (hubs) are driving the evolutionary path and are figuring out the synchronization patterns by absorbing the small clusters. As can be seen in Figure 6 and Figure 7, two giant groups are recognizable after 14 mm displacement. This shows the attractors states in a dynamic system. However, two discriminated clusters are not showing the self-similarity structures within the proper networks, i.e., hubs with high degree nodes are separated from the hubs with low degree nodes. In general, one may overestimate the self-similarity of internal structures of the networks, which means that in the entire steps at least a small branch of fractility can be followed.

The attributed weight distribution, associated with the two-point correlation concept (Fig.7) shows as if the virtual heaviness of edges are increasing, simultaneously, the joint degree distribution is also growing, which indicates the networks are assortative. The distribution of the weights from unveiled hubs also clearly can be followed in Figure 7 while two general discriminated patterns are recognizable. On the contrary, if the patterns of correlation of clustering coefficients are drawn (Fig.8), the eruption of local synchronization is generally closed out after (or at least near) peak point while again during and after dropping shear strength, the variation of local clusters will continue. Especially, at the point near to critical step, the local clusters present much more uniform percolation rather than the other states while at final steps the stable state (or quasi-stable) regime of regional structures is not clear. It is worth stressing the rate of variation of local joint clustering patterns at apparently quasi-steps are much higher than the global patterns, i.e., joint degree distribution. Also, it must be noticed that before peak point the structures of joint triangles density is approximately unchangeable. Then as a conclusion, burst of much dense local hubs is scaled with disclosing of slow fronts spreading.

Following the spectrum of the networks in a collective view (Fig.9) shows a nearly uniform growing trend where a third degree polynomial may be fitted. However, with respect to individual analysis (local analysis) of $c_i - k_i$, a negative trend can be pursued. The spectrum of the networks can be related to three-point correlation concept which expresses the probability of selecting a node with a certain degree, so that it is connected to other two nodes with the definite degrees. The evolution of spectrum of aperture networks in a Euclidean space and using a clustering analysis on the accumulated objects has come out the details of the fracture evolution, either in the mechanical or hydro-mechanical analysis [18-19]. But, in our case, detecting such explicit scaling is difficult. Let us transfer all of the calculated network properties in a variation (rate) space (Fig.10). Depicting the clustering coefficient and mean degree rates, shows a similar trend with the evolution of shear strength, however, after 8 mm displacement the variation of edges and clustering coefficient unravels the different fluctuations.

The negative scaling (for large anisotropy) in $\frac{dc_i}{dt} - \frac{dk_i}{dt}$ space can be expressed by $\frac{dk_i}{dt} \cong -800\frac{dc_i}{dt} + 20$. As it can be followed in Figure 10, the congestion of objects makes a general elliptic which approximately covers all of points where the details of the correlation among two components presents how the expansion and contraction of patterns fall into the final attractors (Fig.10). Thus, such emerged patterns related to the two-point correlation of variation rate of edges and rate of clustering coefficient are proposing a certain core in each time step so that the absorbing of objects within a "black hole" at residual part is much more obvious rather than other states. With definition of anisotropy by $S = \sigma(\frac{d \prec k \succ}{dt}) \Big/ \sigma(\frac{dC}{dt})$ ($\sigma$ is standard deviation), the rate changes of profiles in a new space and with reference to the pre and post peak behaviours are obtained (Fig.11). Transferring from interlocking step to Coulomb threshold level is accompanying with the maximum anisotropy (Fig.11b) and immediate dropping and then starting to fluctuate until reaching to a uniform decline. The fluctuation of anisotropy from 2mm to 13 mm may be associated with the stick-slip behaviour of the rock joint as the main reason of shallow earthquakes [57-58]. It should be noticed that the results of the later new space is completely matching with the analysis of joint degree and joint clustering distribution. In Figure 11.a, we have illustrated a new variable with regard to durability and entropy of the system, $\frac{dC}{dt}\times\frac{d\prec k\succ}{dt}$. In fact with definition of such parameter the fluctuation in anisotropy is filtered while initiating the post stick-slip behavior is scaled with the minus or zero variation of the parameter. In [59-61], we analyzed the sub-graph structures and frequencies over parallel and perpendicular aperture networks. Also, a directed network based on contact strings and preferentiality of possible energy flow in rupture tips has been introduced. We also, inspected the synchronization of strings using a Kuromoto model [59].

## 5. Conclusions

In this study, we presented a special type of complex aperture network based on correlation measures. The main purpose of the study was to make a connection between the apparent mechanical behavior of a rock joint and the characterized network. The incorporation of the correlation of apertures and the evaluation of continuously changing contact areas (i.e. growth of aperture) within the networks showed the effects of structural complexity on the evolution path of a rock joint. Our results showed that the main characteristics of aperture networks are related to the shear strength behavior of a rock joint. The residual shear strength corresponded to the formation of giant groups of nodes in the networks. In addition, based on the joint correlation upon edges and triangles, the pre-peak and post peak behaviour of a rock joint under shear were analyzed. Our results may be used as an approach to insert the complex aperture networks into the surface growth methods or general understanding of the conditions for a sudden movement (shock) in a fault.

Figures

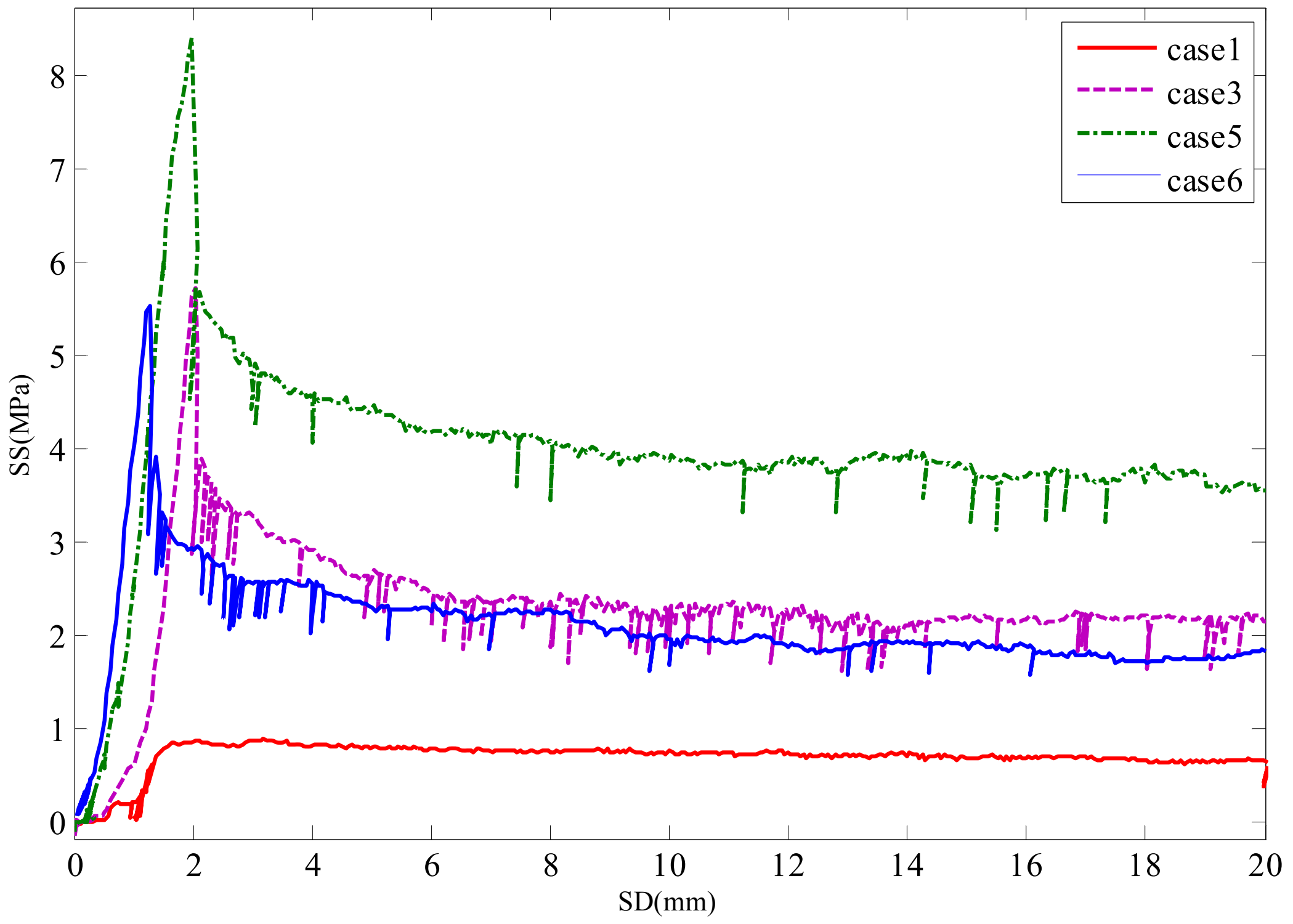

FIGURE 1. Variation of shear strength for different cases (normal stresses for case1:1mpa, case3: 3 MPa, case5: 5 MPa and case6: 3 MPa (without control of upper shear box) [46].

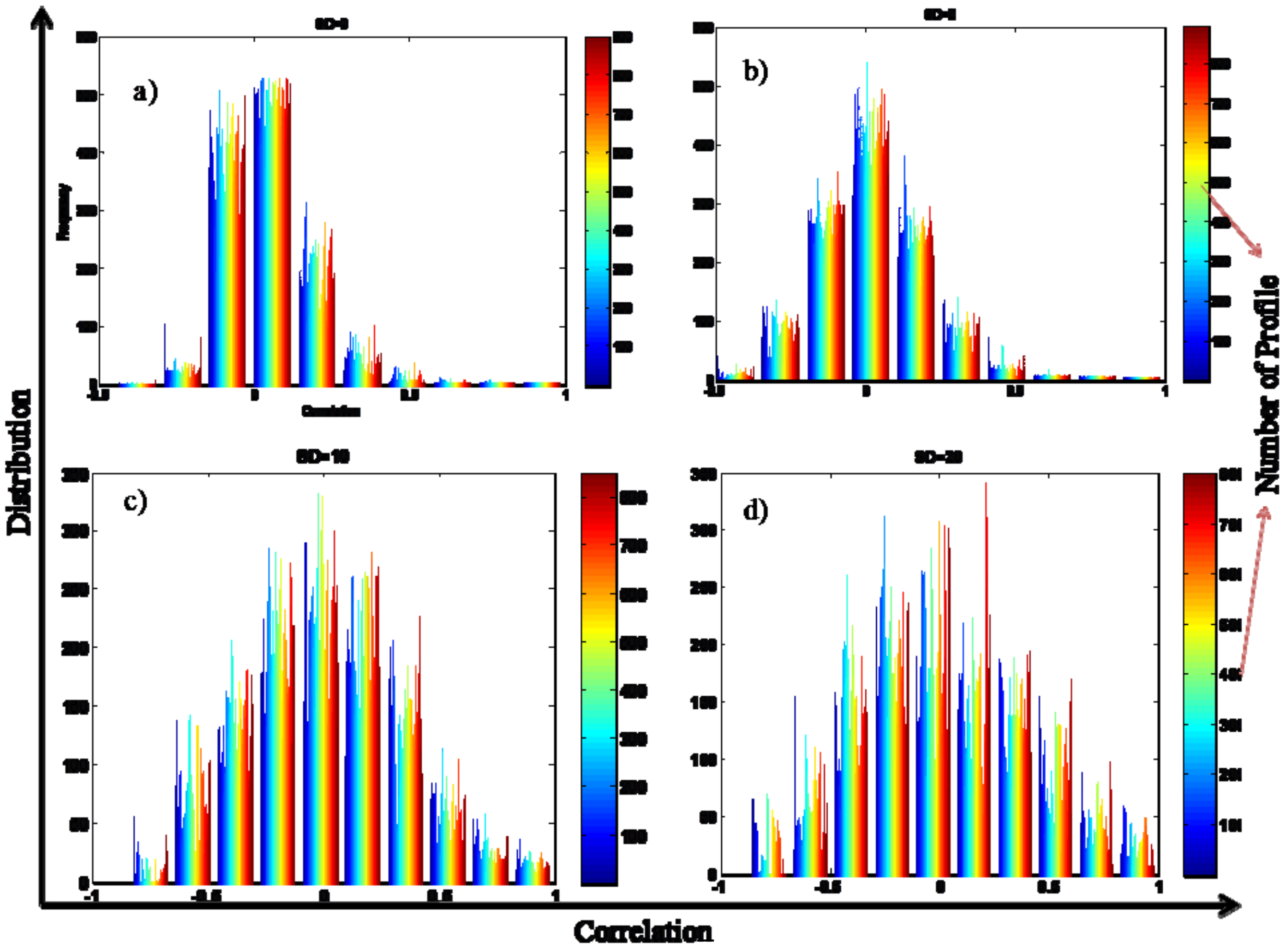

FIGURE 2. Evolution of correlation values of aperture profiles at shear displacement s: 0,2,10 and 20 mm.

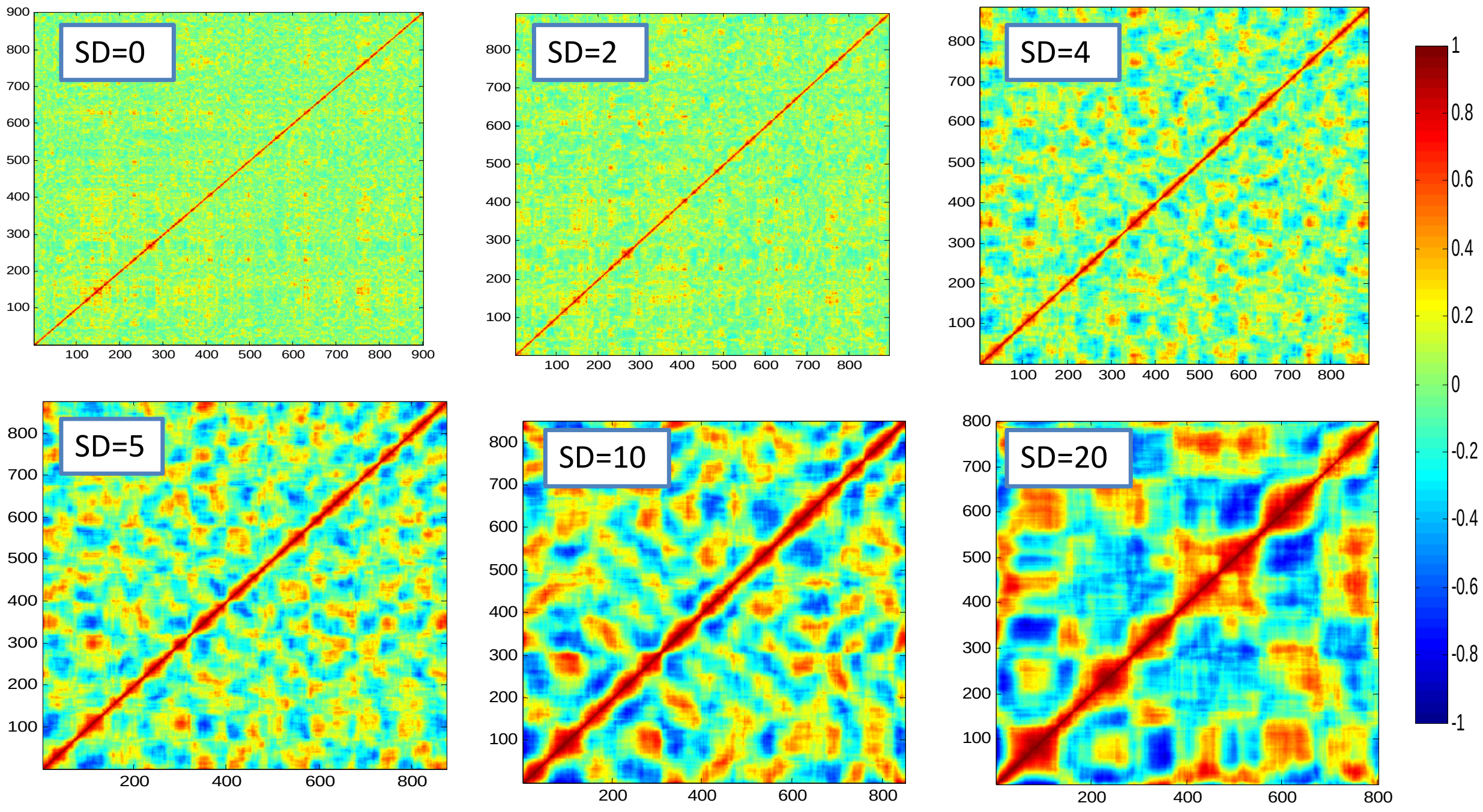

FIGURE 3. Correlation patterns throughout the shear displacements

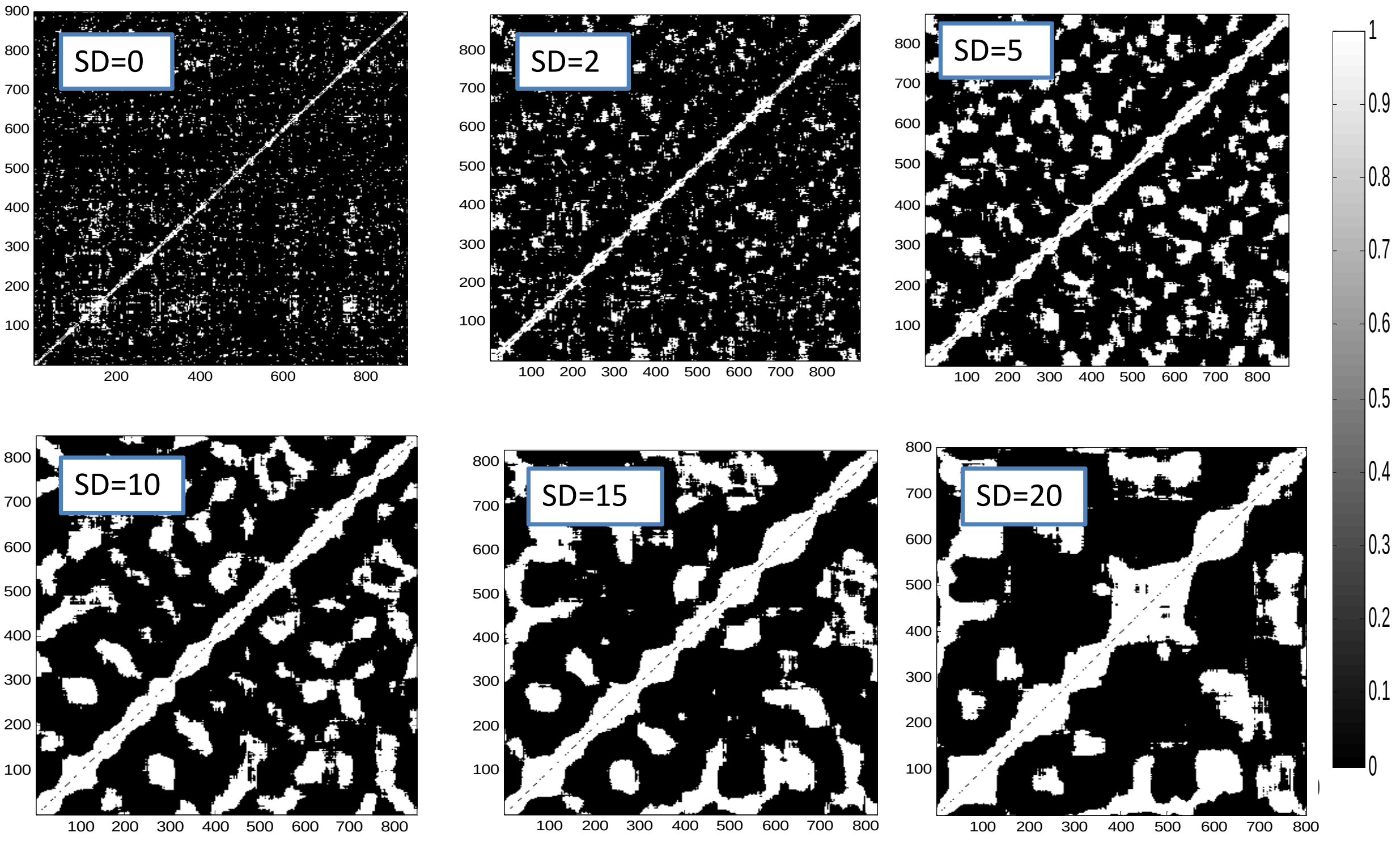

FIGURE 4. Visualization of adjacency matrix for the achieved networks

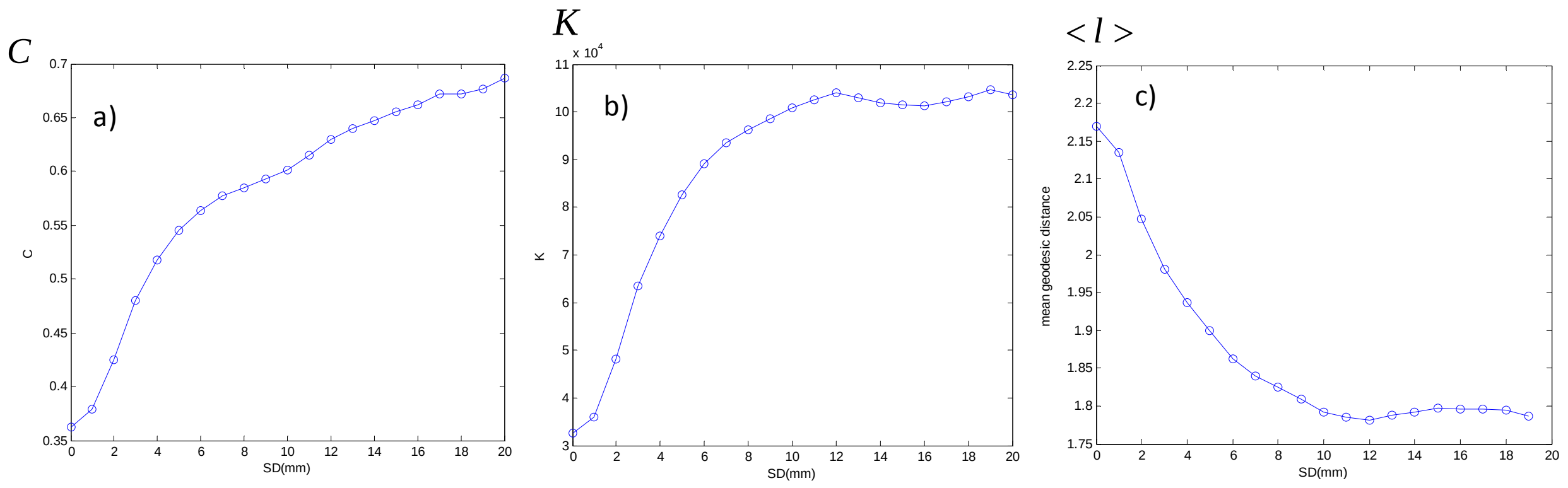

FIGURE 5. a) Clustering coefficient-Shear Displacement (SD), b) Number of edges-SD and c) Average path length-SD

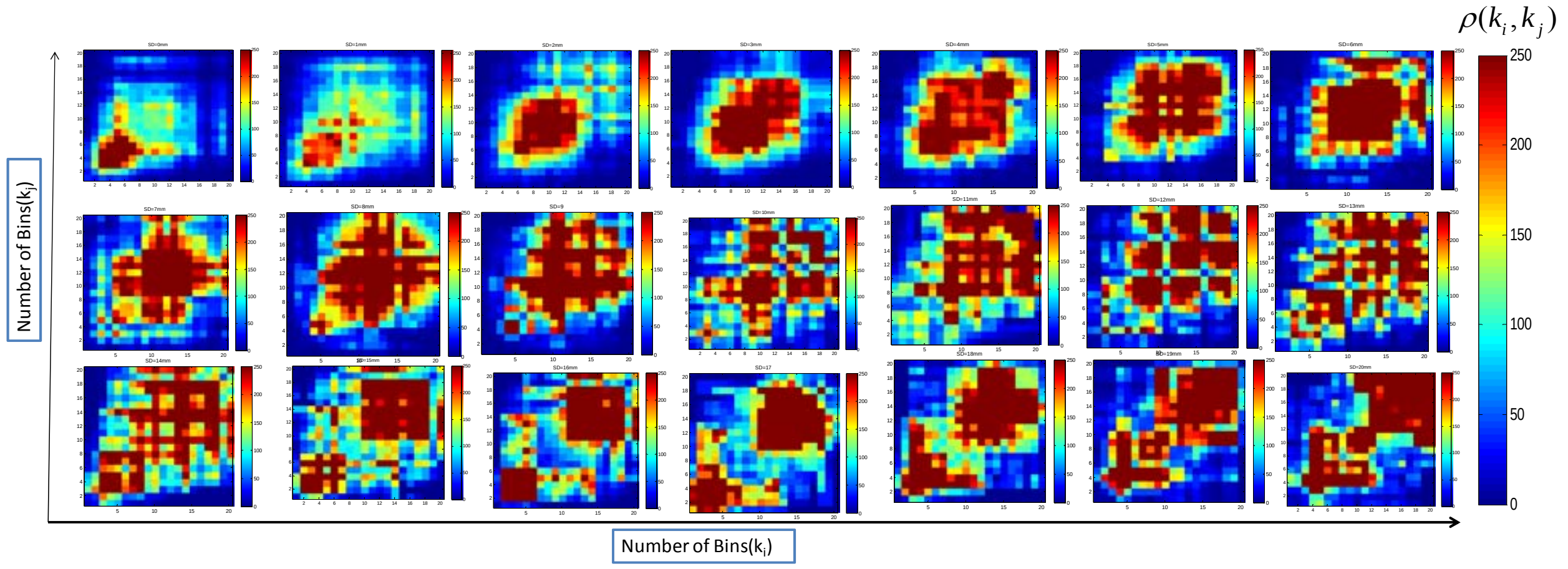

FIGURE 6. Joint degree distribution from SD=0 to SD=20 mm (Top left-first row is SD=0 and Top right -first row is 6mm shear slip )

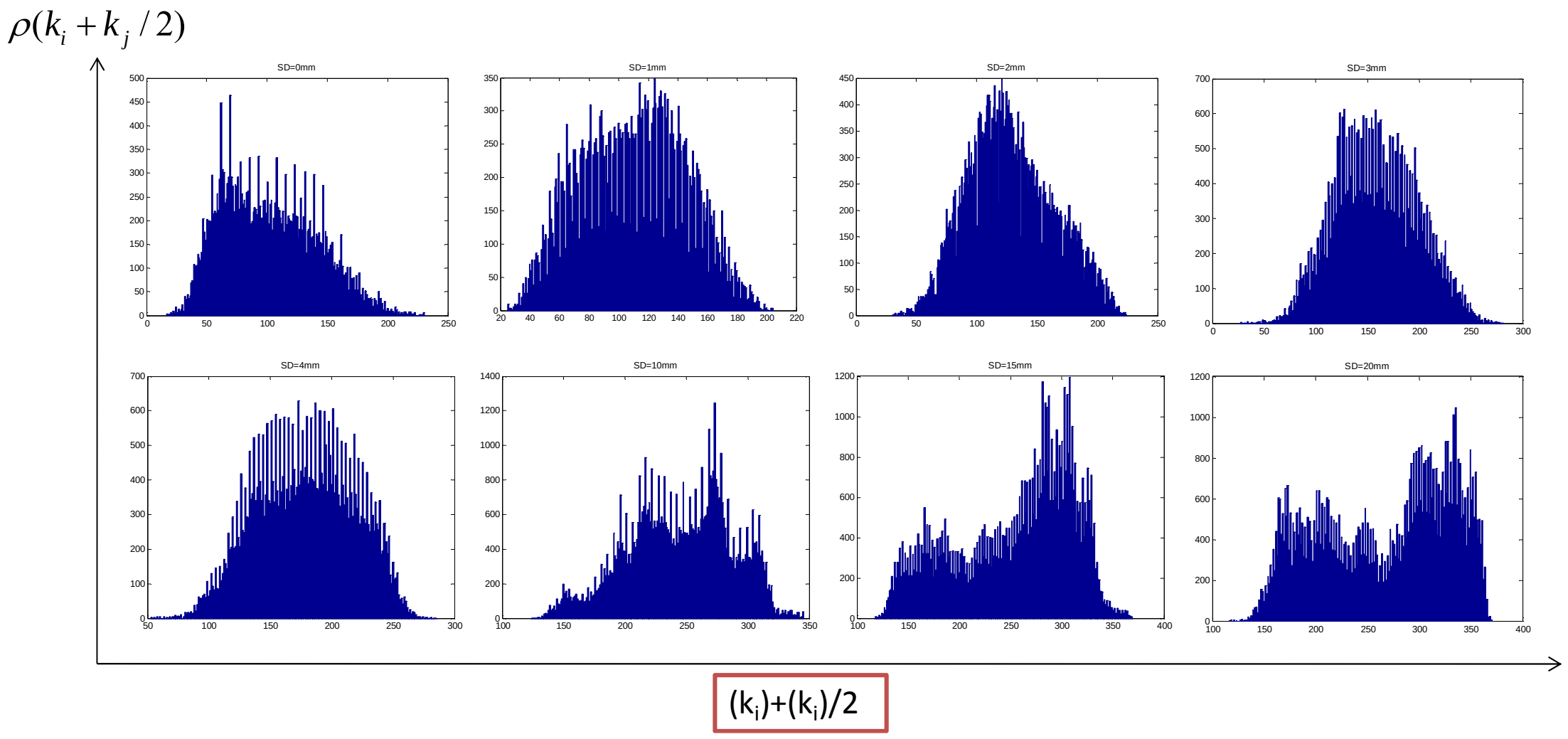

FIGURE 7. Attributed weight distribution of links related to joint degree distribution (for SD=0-4,10,15 and 20 mm)

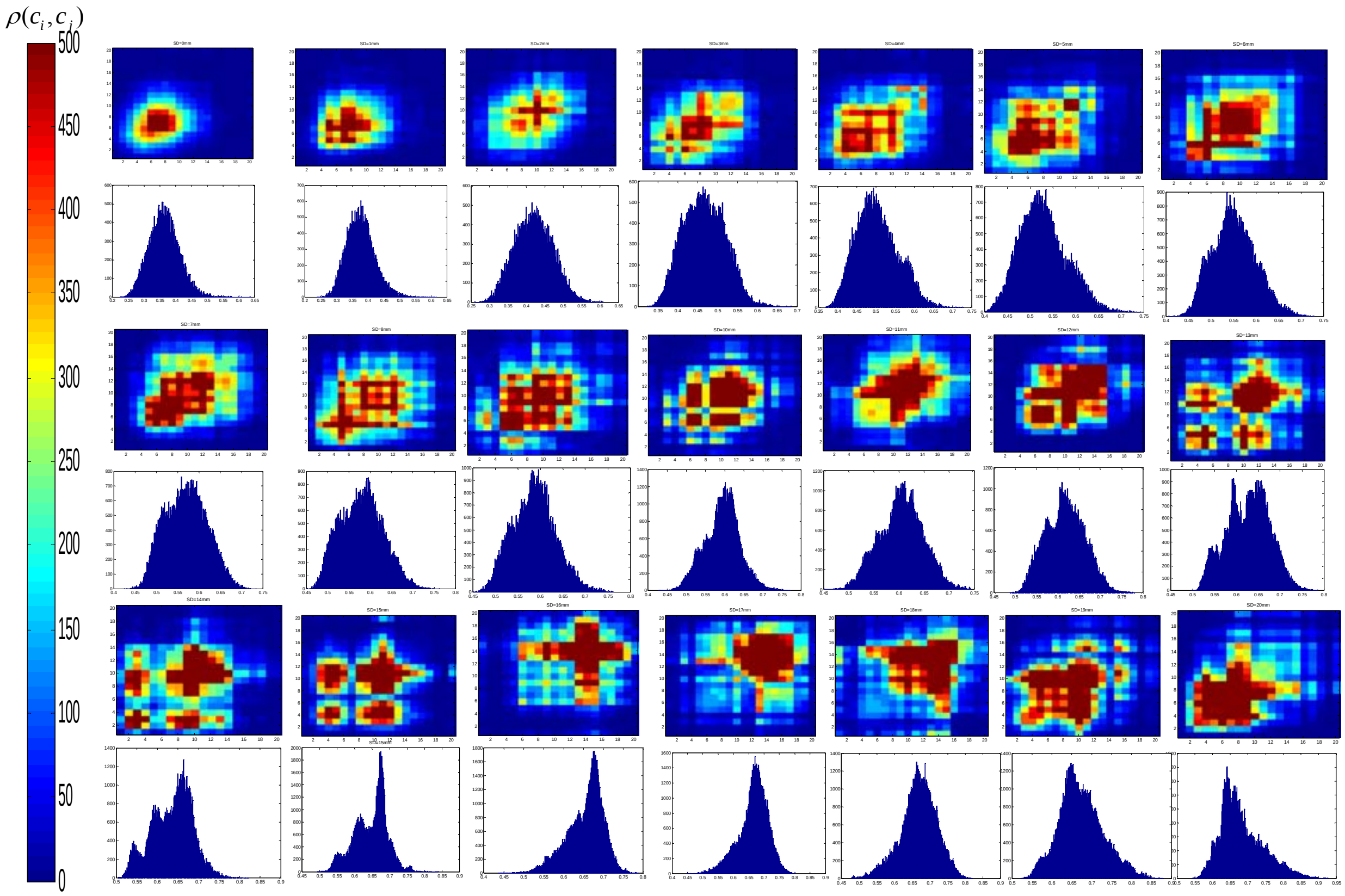

FIGURE 8. Joint clustering coefficient distribution plus attributed weight histograms based on averages of triangles connected to a link (sequence of figures are as well as figure 6).

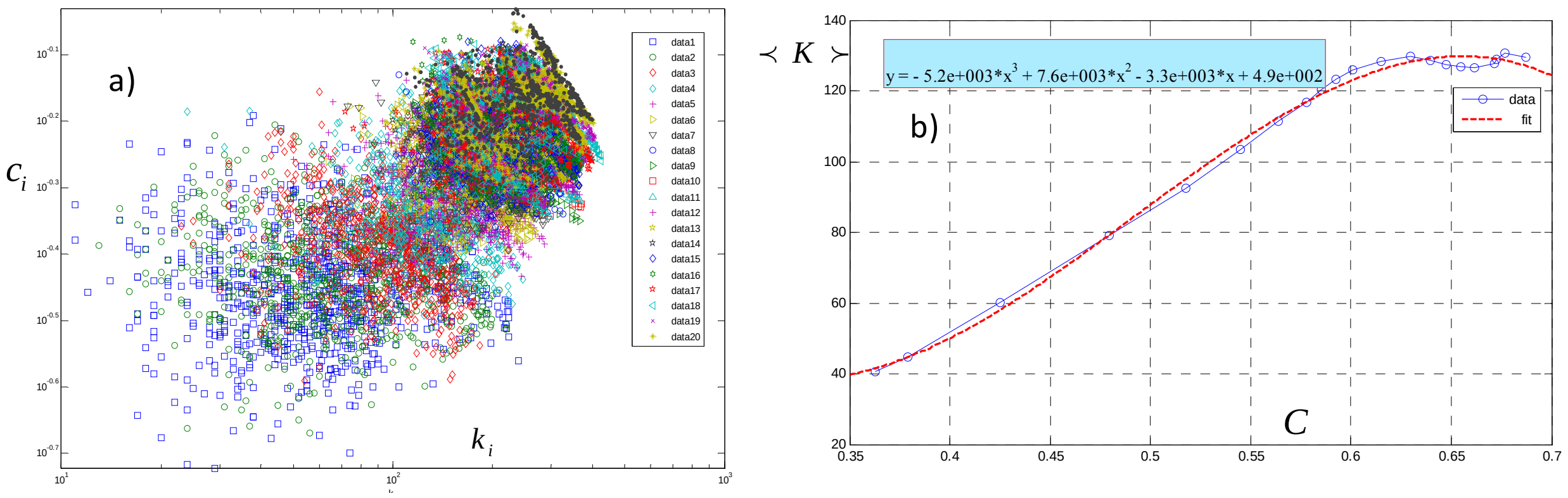

FIGURE 9. a) Spectrum of complex aperture networks ($c_i$ - $k_i$)and b) Evolution of mean degree of node against clustering coefficient and fitness of a polynomial function

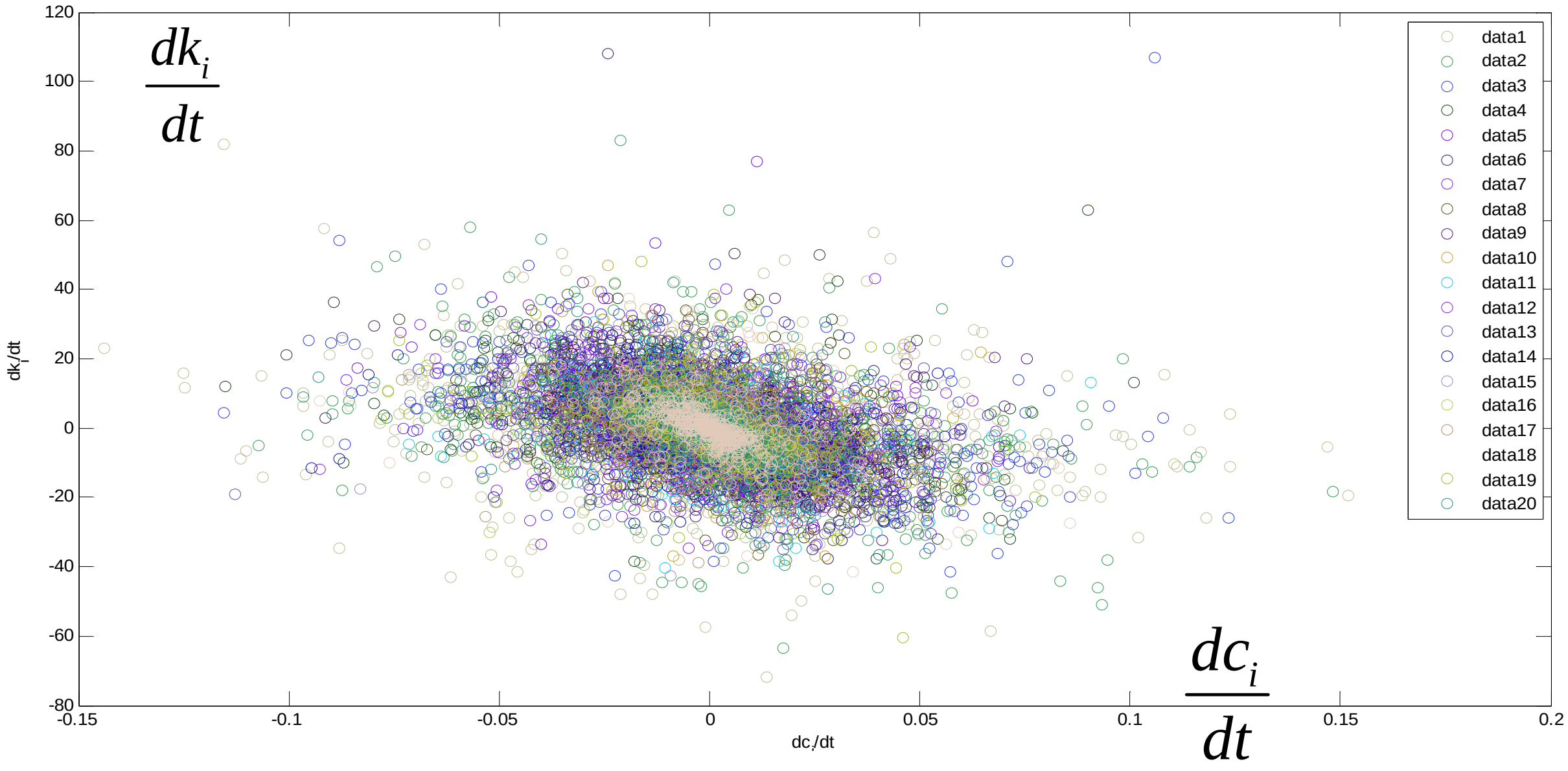

FIGURE 10. Data accumulation in $\frac{dk_i}{dt} - \frac{dc_i}{dt}$ space with respect to shear displacements (data1to data 20 are related to shear displacements from 0 to20 mm ).

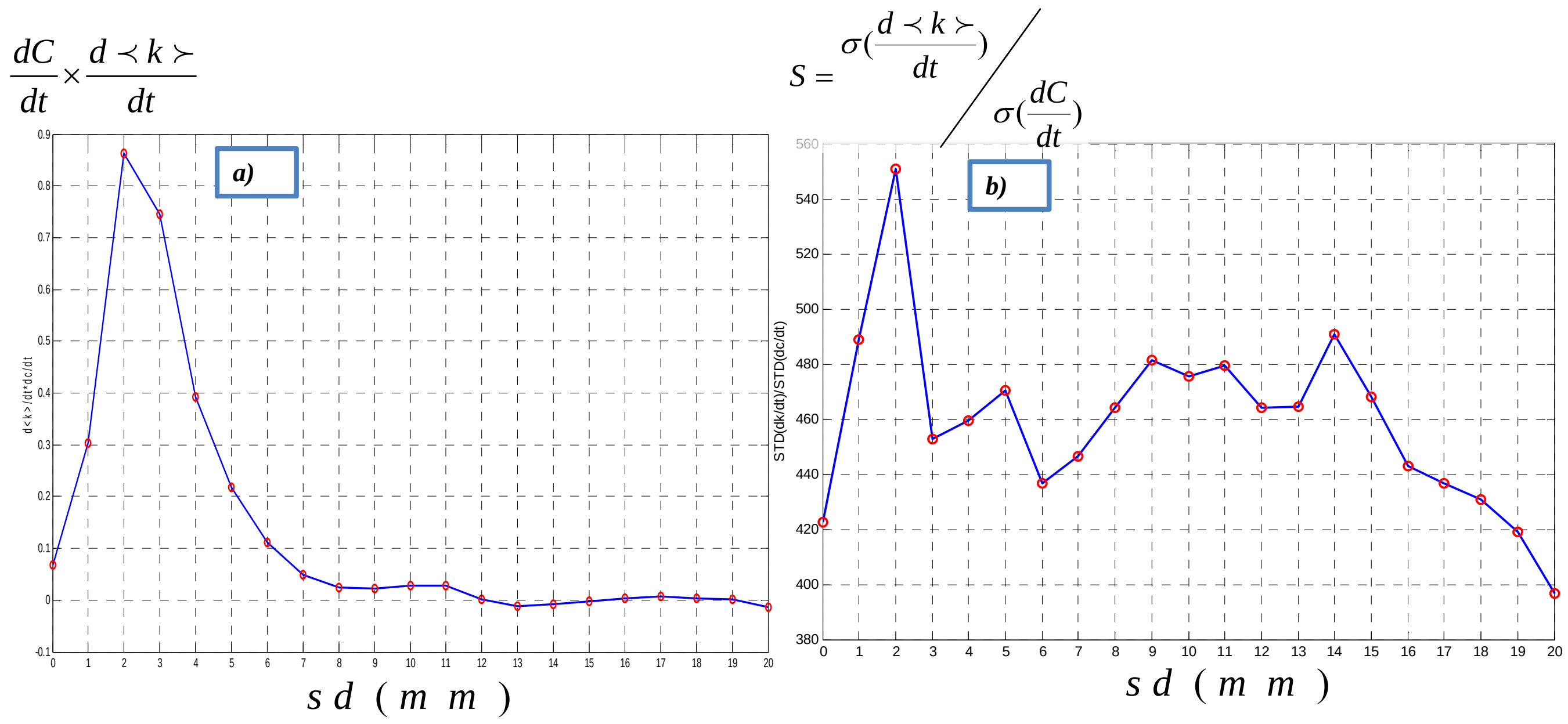

FIGURE 11.a) Variation of $\frac{dC}{dt}\times\frac{d<k>}{dt}$ with shear displacements and b) Anisotropy evolution at the rate of spectrum (networks) space